\documentclass[twocolumn,prx,superscriptaddress,longbibliography]{revtex4-1} 
\usepackage[utf8]{inputenc}
\usepackage[english]{babel}
\usepackage[T1]{fontenc}
\usepackage{amsmath}

\usepackage{tikz}
\usepackage{lipsum}

\usepackage[utf8]{inputenc}
\usepackage{amssymb}
\usepackage{amsmath}
\usepackage{color}
\usepackage[colorlinks=true,urlcolor=blue,breaklinks,citecolor=blue]{hyperref}
\usepackage{graphicx}
\usepackage{braket}
\usepackage{xcolor}
\usepackage{dcolumn,bm}
\usepackage[capitalize]{cleveref}
\usepackage{float}
\usepackage{here}
\usepackage{hyperref}
\usepackage{mathtools}

\DeclarePairedDelimiter\floor{\lfloor}{\rfloor}

\usepackage{soul}

\usepackage{xcolor}
\usepackage{listings}
\definecolor{mygray}{rgb}{0.8,0.8,0.8}
\usepackage{realboxes}

\newcommand{\code}[1]{\lstinline|#1|}

\usepackage[draft]{changes}
\usepackage[justification=justified, format=plain]{caption}
\usepackage{subcaption}

\usepackage{ragged2e}
\DeclareCaptionJustification{justified}{\justifying}
\begin{document}
\title{Variational Quantum Anomaly Detection: Unsupervised mapping of phase diagrams on a physical quantum computer}
\author{Korbinian Kottmann}
\affiliation{ICFO - Institut de Ciencies Fotoniques, The Barcelona Institute of Science and Technology, Av. Carl Friedrich Gauss 3, 08860 Castelldefels (Barcelona), Spain}
\author{Friederike Metz}
\affiliation{Quantum Systems Unit, Okinawa Institute of Science and Technology Graduate University, 1919-1 Tancha, Onna, Okinawa 904-0495, Japan}
\author{Joana Fraxanet}
\affiliation{ICFO - Institut de Ciencies Fotoniques, The Barcelona Institute of Science and Technology, Av. Carl Friedrich Gauss 3, 08860 Castelldefels (Barcelona), Spain}
\author{Niccol\`o Baldelli}
\affiliation{ICFO - Institut de Ciencies Fotoniques, The Barcelona Institute of Science and Technology, Av. Carl Friedrich Gauss 3, 08860 Castelldefels (Barcelona), Spain}

\date{\today}

\begin{abstract}
One of the most promising applications of quantum computing is simulating quantum many-body systems. However, there is still a need for methods to efficiently investigate these systems in a native way, capturing their full complexity. Here, we propose variational quantum anomaly detection, an unsupervised quantum machine learning algorithm to analyze quantum data from quantum simulation. The algorithm is used to extract the phase diagram of a system with no prior physical knowledge and can be performed end-to-end on the same quantum device where the system is simulated on. We showcase its capabilities by mapping out the phase diagram of the one-dimensional extended Bose Hubbard model with dimerized hoppings, which exhibits a symmetry protected topological phase. Further, we show that it can be used with readily accessible devices today by performing the algorithm on a real quantum computer.
\end{abstract}

\maketitle

\section{Introduction}
Since the discovery of Shor's algorithm \cite{Shor1994}, there have been many attempts to leverage the power of quantum computers to outperform classical computers \cite{nielsen_chuang_2010}. One of the most promising near- and far-term applications is simulating quantum many-body systems.
Universal quantum simulation algorithms like phase estimation \cite{Kitaev1995, nielsen_chuang_2010}, or adiabatic quantum computation \cite{Farhi2000} require quantum fault-tolerance \cite{Aharonov1999}, that is, the ability to reliably correct errors that occur during the quantum computation. With recent experimental advancements in quantum error correction \cite{Semeghini2021,Satzinger2021}, there is hope that quantum fault-tolerance can one day be reached.
Currently, we only have access to noisy intermediate-scale quantum (NISQ) devices \cite{Preskill2018}. There are several proposals for algorithms on these devices \cite{Bendetti2019,bharti2021noisy} like the variational quantum eigensolver (VQE) \cite{Peruzzo2014}, the quantum approximate optimization algorithm (QAOA) \cite{Farhi2014}, or the quantum autoencoder \cite{Romero2017,Prieto2021}, that employ parameterized circuits which are optimized through a classical feedback loop, typically with gradient based methods \cite{McClean2015, Stokes2019, Cerezo2020_variational}. These approaches can suffer from so-called Barren Plateaus, the phenomenon of an exponentially vanishing gradient of the loss function \cite{McClean2018}. In practice, this issue occurs for large systems but can be avoided by using shallow circuits and local cost functions \cite{Cerezo2020}. While gradient-free optimization does not solve the Barren Plateau problem \cite{Arrasmith2021}, other proposals give hope for large-scale and deep variational quantum circuits \cite{Pesah2021,Volkoff2021}.

With the rise of deep learning in the 2010s, the term \textit{quantum machine learning} was mostly used to refer to leveraging quantum computers for linear algebra tasks such as matrix inversion in sub-polynomial time via the Harrow-Hassidim-Lloyd algorithm \cite{Harrow2008, Biamonte2017}. One famous use-case was the quantum recommendation system algorithm with an exponential quantum speed-up at the time \cite{Kerenidis2016}, which inspired classical analogs of the algorithm with the same, sub-polynomial, complexity (termed as \textit{quantum-inspired} machine learning algorithms) \cite{Tang2018}. Today, quantum machine learning refers to using quantum circuits as neural networks \cite{Perez-Salinas2020}, or kernel functions \cite{Schuld2021} to perform classical machine learning tasks like supervised learning \cite{Farhi2018,Rebentrost2013}. There are cases, where quantum models have provable advantages over classical models \cite{Liu2020}, but it has been argued that these instances are special cases and no quantum speed up is to be expected for quantum machine learning with classical data \cite{Kubler2021}.

On the other hand, applying classical machine learning to quantum physics has been a great success story \cite{Carleo2019}, most prominently for the classification and mapping of phase diagrams \cite{Carrasquilla2016,VanNieuwenburg2016,Huembeli2018}. These methods rely on classical data and are therefore restricted by the available classical simulation methods. With physical devices surpassing system sizes that are classically tractable \cite{Arute2019}, there is need for methods to investigate physical quantum states with quantum computers.

In this paper, we propose a quantum machine learning algorithm for \textit{quantum data}. The data are ground states of quantum many-body systems that are prepared by a quantum simulation subroutine and serve as the input for \textit{Variational Quantum Anomaly Detection} (VQAD). Our quantum anomaly detection scheme belongs to the category of variational quantum algorithms where the circuit \textit{learns} characteristic features of the input state \footnote{The term \textit{learning} is commonly used in (quantum) machine learning and data-driven problem solving to refer to data-specific optimization.}. This can in principle be leveraged for obtaining physical insights of the system from training \cite{Iten2018} and is in contrast to previous proposals that are based on kernel methods (one-class support vector machines) \cite{Liu2017,Liang2019}. In the present study, we use it to map out an unknown phase diagram of a system without requiring knowledge about the order parameter or the number and location of the different phases.

In anomaly detection, the task is to differentiate \textit{normal} data from \textit{anomalous} data, opposed to supervised learning tasks, where a fixed set of classes with labels for training are differentiated. On the other hand, the task of anomaly detection requires an \textit{anomaly syndrome}, i.e., an observable that is trained to be of a certain value (typically $0$) when \textit{normal} data is input, and be significantly larger for \textit{anomalous} data it is tested on. In classical machine learning, anomaly detection has already been used to extract phase diagrams in an unsupervised fashion from simulated and experimental data \cite{Kottmann2020,Kaming2021, Kottmann2021}. VQAD allows us to perform anomaly detection directly \textit{on} a quantum computer, and, with programmable devices readily available, we demonstrate it experimentally on a real device.

\section{Proposal}

\begin{figure}
    \includegraphics[width=.48\textwidth]{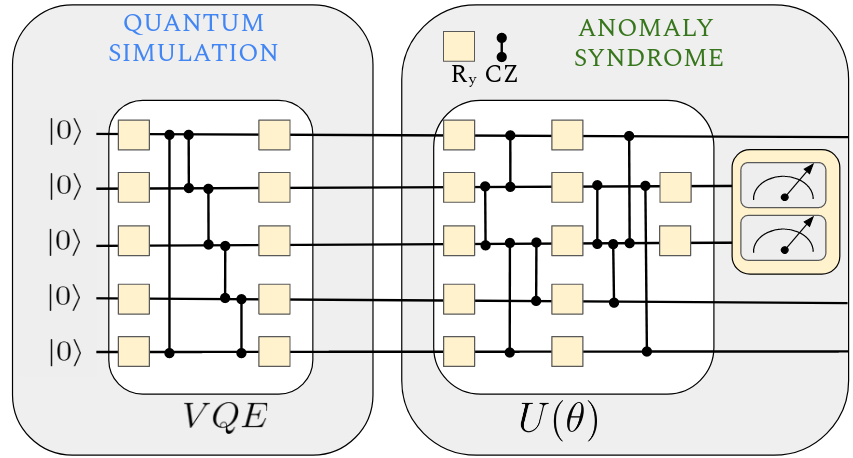}
    \caption{Overview of our proposal. First, the quantum states are prepared via VQE. Then, they are processed through the anomaly syndrome, consisting of a parameterized unitary $U(\theta)$ and a measurement of a subset of qubits, referred to as trash qubits. $R_y$ indicates a parameterized y-axis rotation and $CZ$ a (fixed) controlled-z gate.}
    \label{fig:overview}
\end{figure}

The task of detecting anomalies in ground states of quantum many-body Hamiltonians can be loosely divided into two sub tasks:~Preparing the ground state for specific Hamiltonian parameters, and computing an anomaly syndrome indicating whether the state corresponds to a \textit{normal} example or an anomaly. An overview of our proposed algorithm is shown in \cref{fig:overview}. 
The problem of state preparation on quantum computers is one of ongoing research, and in principle, one can use any state preparation subroutine for preparing the ground state. Here, we choose the Variational Quantum Eigensolver (VQE) as it has the lowest hardware requirements while achieving reliable results on current devices \cite{Peruzzo2014,Kandala2017}. The VQE algorithm iteratively minimizes the expectation value of a Hamiltonian with the ansatz circuit to find the ground state by optimizing the parameters of the circuit via a quantum-classical feedback loop. We choose a minimal ansatz as depicted in \cref{fig:overview} that is sufficient for simulating the Ising Hamiltonian discussed in Sec.~\ref{sec:exp}. A shallow ansatz allows us to run both, the quantum simulation, and the quantum anomaly detection on real noisy devices. For more complex systems, the problem of finding a suitable hardware efficient ansatz can be addressed for example by the adaptive VQE algorithm \cite{2019adaptvqe}. In this work we employed the VQE implementation provided by the Qiskit library \cite{Qiskit} and optimized it using simultaneous perturbation stochastic approximation (SPSA) \cite{Spall1998}. For all technical details we refer to App.~\ref{app}.

\begin{figure}
    \centering
    \includegraphics[width=\columnwidth]{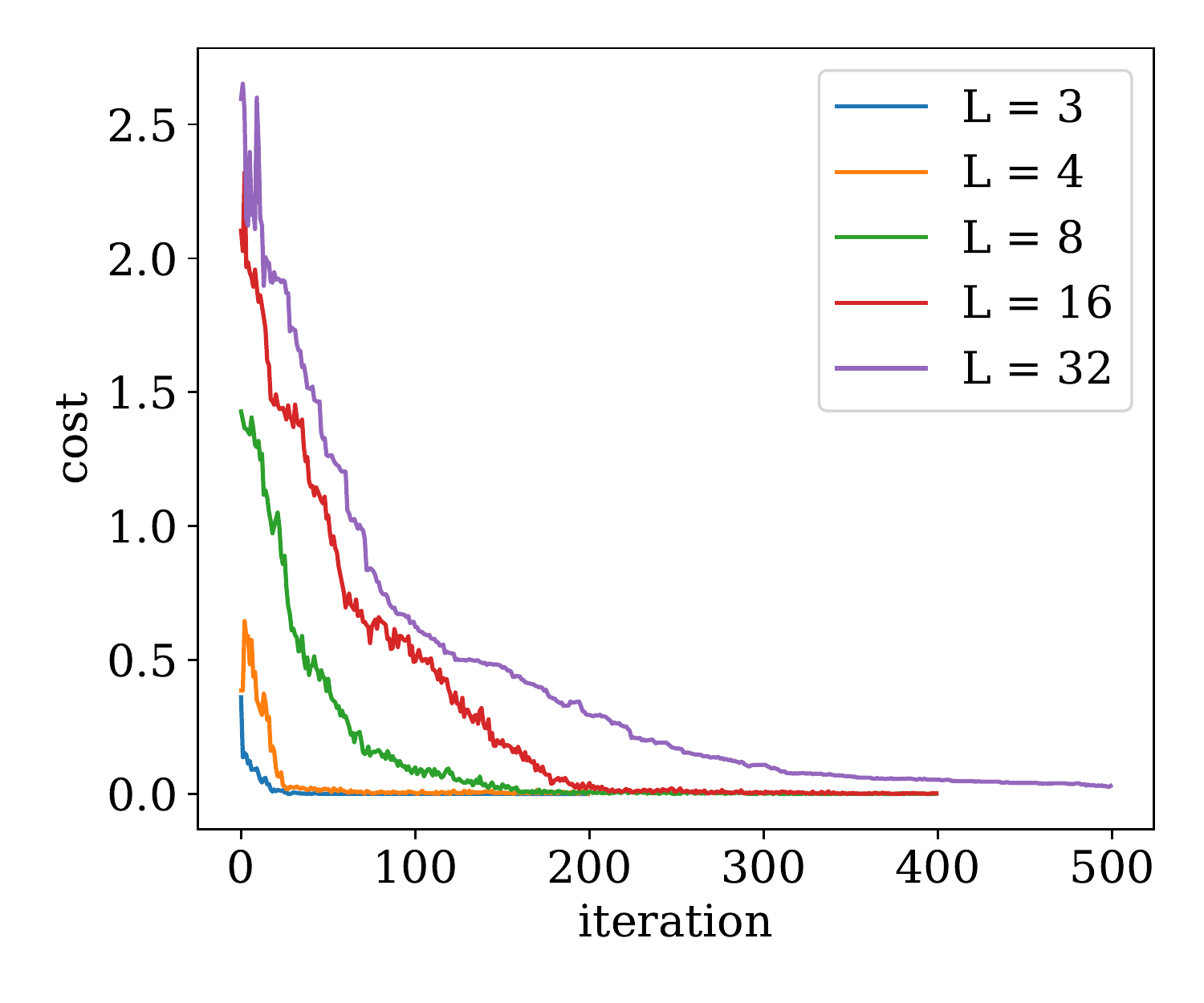}
    \caption{Scaling of the training cost of the anomaly syndrome ansatz. Successful training of the proposed anomaly syndrome ansatz for $L \in \{3,4,8,16,32\}$ corresponding to $n_t \in \{1,2,3,4,5\}$ trash qubits (and therefore $n_t$ circuit layers). The result for $L=32$ was obtained through MPS simulations with a maximal bond dimension of BD = $100$. We used $1000$ shots per evaluation and achieve a perfect cost value of $0.00$ for all system sizes, however the run for $L=32$ shown here finished at $0.03$.
    }
    \label{fig:ansatz_scaling}
\end{figure}

Once the ground state is prepared on the quantum device a subsequent circuit serves as the anomaly syndrome. Our circuit ansatz is inspired by the recently proposed quantum auto-encoder, which similar to its classical counterpart can be used for compression of classical and quantum data \cite{Romero2017,Prieto2021}. It is composed of several layers each consisting of parameterized single qubit y-rotations and controlled-z gates. After the final layer a predefined number $n_t$ of \textit{trash} qubits is measured in the computational basis. The objective is to decouple the trash qubits from the rest of the system, effectively compressing the original ground state into a smaller number of qubits. The circuit parameters are then optimized to faithfully compress states that are considered \textit{normal}. However, when the optimized circuit is tested on anomalous states not seen during training, it is expected that the circuit fails to decouple the trash qubits from the rest of the system. To quantify the degree of decoupling we use the Hamming distance $d_{H}$ of the trash qubit measurement outcomes to the $|0\rangle^{\otimes n_{t}}$ state, i.e., the number of 1s in a bit-string of measurement outcomes \cite{Prieto2021}. The cost function $C$ can then be defined as the Hamming distance averaged over several circuit evaluations $C = 1/N\sum_i^N d_{Hi}$, where $N$ is the number of performed measurements or shots. The cost function can also be rewritten in terms of expectation values of local Pauli-z operators $Z_j$
\begin{equation}
	C = \frac 1 N \sum_{i=1}^N d_{Hi} = \frac{1}{2} \sum_{j=1}^{n_{t}}\left(1-\left\langle Z_{j}\right\rangle\right) .
\end{equation}
The VQAD circuit achieves perfect compression if the trash qubits are fully disentangled from the remaining qubits and mapped into the pure $|0\rangle^{\otimes n_{t}}$ state resulting in a cost equal to zero.

The specific circuit ansatz for the anomaly syndrome is shown in \cref{fig:overview} for the case of $n_t=2$ trash qubits. Each layer of the circuit starts with parameterized single-qubit y-rotations applied to every qubit followed by a sequence of entangling controlled-z gates. The currently available NISQ devices are inherently noisy and the computations are subject to gate errors. To minimize the number of two-qubit gates we apply the controlled-z gates only between trash qubits and non-trash-qubits as well as between trash qubits themselves instead of an all-to-all entangling map \cite{Prieto2021}. This entangling map is physically motivated as the goal of the circuit is to disentangle the trash qubits from the rest, with the trash qubits resulting in the $|0\rangle^{\otimes n_{t}}$ state. In a single layer each non-trash qubit will be coupled to exactly one trash qubit. This entangling scheme is repeated in the subsequent layers until every non-trash qubit has been coupled to each trash qubit exactly once, i.e. the number of layers of the circuit is equal to $n_t$. After the final layer, additional single-qubit y-rotations act on the trash qubits before they are measured.

Barren Plateaus are the fundamental obstacle prohibiting training of variational circuits with increasing numbers of qubits \cite{McClean2018}. It was previously shown that using local cost functions and circuits featuring a number of layers scaling at most logarithmically in the system size can prevent the occurrence of Barren Plateaus \cite{Cerezo2020}. Additionally for realistic devices, gate errors lead to decoherence, making quantum simulation on real devices a challenging task even for small systems and low depths \cite{CerveraLierta2018}. The former calls for a minimal number of layers while the latter calls for a minimal number of gates overall. Therefore, we seek a minimal solution for our variational circuit that we want to implement on a readily available NISQ-era quantum computer. On the other hand, it is desirable to have an ansatz as general as possible to be able to capture a wide range of problems (see \textit{circuit complexity} \cite{Bernstein1997,Brandao2019}).

For the anomaly syndrome in this paper, we propose an ansatz that aims at compromising between being general enough to compress the ground states of the investigated systems while still being trainable. One way to make our circuit scalable for larger systems is to choose the number of trash qubits $n_t = \floor*{\log_2 L}$, where $L$ is the total number of qubits. Together with the fact that our cost function is composed of only local operators, the training is expected to not suffer from Barren Plateaus. We empirically confirm successful trainability, i.e., achieving a cost of $0.0$ for ground states of the systems discussed later in the manuscript, for $L \in \{3,4,8,16, 32\}$, corresponding to $n_t \in \{1,2,3,4,5\}$, respectively. In \cref{fig:ansatz_scaling}, a ground state of the Ising model \cref{IsingHam} at $g_x, g_z, J = (0.3,0,1)$ is taken as a realistic example and we can confirm successful trainability in all cases.

Note that in principle the trash qubits can be placed anywhere in the circuit, however, when performing computations on a real quantum device it proved advantageous to explicitly take the qubit connectivity structure of the device into account in order to reduce the number of required SWAP operations. Specifically here, we placed the trash qubits in the middle of the IBMQ devices.



\begin{figure*}
	\centering
	\includegraphics[width=\textwidth]{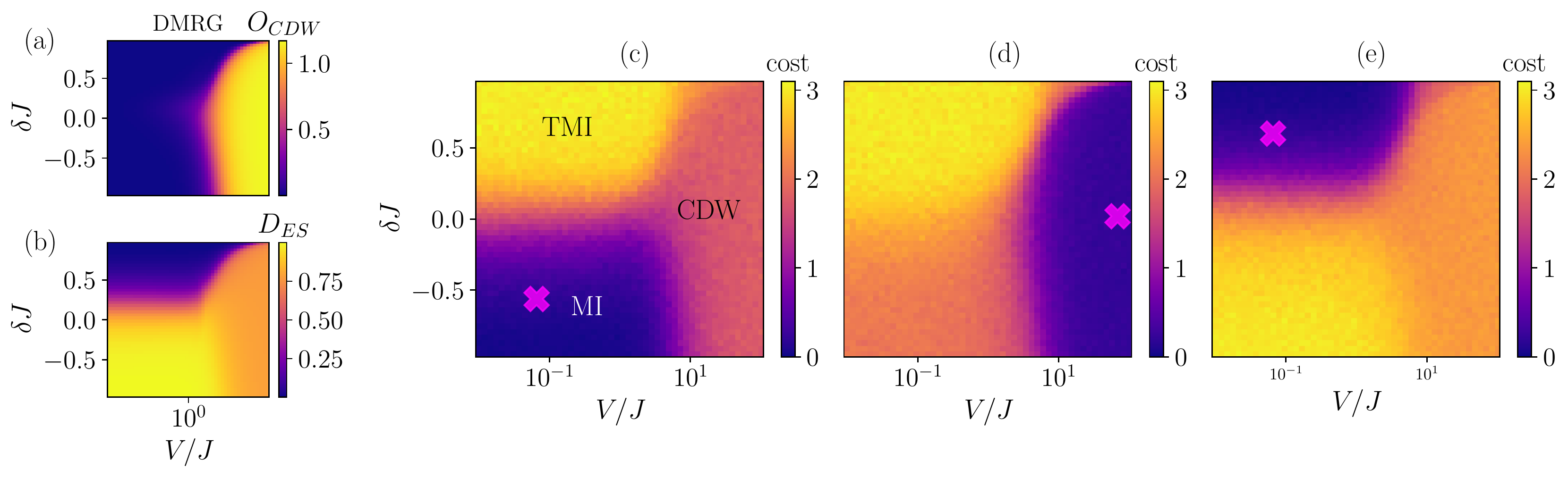}
	\caption{(a)-(b) Phase diagram of the DEBHM from Eq.~\eqref{eq:ham_bh} using (a) the order parameter $O_{CDW}$ defined in Eq.~\eqref{eq:OCDW}, and (b) the degeneracy of the entanglement spectrum, $D_{ES}$, defined in Eq.~\eqref{D_ES}. The results were obtained from DMRG simulations for a system of length $L=12$ at half filling $\bar{n} = 0.5$. We fix the maximum bond dimension $BD = 50$ and the maximum number of bosons per site to $n_0 = 1$. (c)-(e) Cost/anomaly syndrome of a VQAD trained on a single ground state (indicated by a cross) of the $L=12$ DEBHM using $n_t=6$ trash qubits in the (c) MI phase, (d) CDW phase, and (e) TMI phase. The cost at each data point is the Hamming distance averaged over 1000 measurement shots using an ideal quantum device simulator.}
	\label{fig:bh}
\end{figure*}

The training and inference procedure is identical to the classical anomaly detection schemes for mapping out phase diagrams \cite{Kottmann2020}. In the first step, one randomly chooses a training region in the phase diagram that represents \textit{normal} data, which is an arbitrary definition. Note that no prior knowledge about the phase diagram is therefore required. The circuit representing the anomaly syndrome is then trained on ground states of the training region, and tested on the whole phase diagram. States in the same phase as the training data are \textit{normal} and can be disentangled, leading to a low cost. \textit{Anomalous} states can be inferred through an increase in the cost function signaling that the corresponding ground state cannot be disentangled by the optimized circuit. From the resultant cost profile, we can deduce the phase boundary between the phase the circuit has been trained on and any other phases in the diagram. 
This procedure is then repeated by training in the anomalous region from the previous iteration until all phase boundaries are found. An example is provided in \cref{fig:bh}.

Anomaly detection is a semi-supervised learning task. The setting is typically that one is provided with one class of data that is well known, \textit{normal} data, and aims at finding outliers of that distribution, \textit{anomalous} data. An archetypical example is credit card fraud where a big database of normal transactions is provided and one aims at finding fraudulent ones. We consider anomaly detection semi-supervised as labeled data (x, “normal”) is provided for training while (x, “anomalous”) is to be inferred. Here, however, we arbitrarily define (x, “normal”) and iteratively find the different classes (phases of matter). The definition of (x, “normal”) is arbitrary and does not necessitate prior knowledge. Furthermore, it is merely a means to an end to find the different classes. In that sense, the way anomaly detection is used to map out the phase diagram can be regarded as an unsupervised learning method.

Note that in previous works, where the same task has been tackled with classical machine learning techniques, it has been shown that a single ground state was sufficient to successfully train the model \cite{Kottmann2021}. This feature stems from the fact that ground states within the same phase share similar properties and there is very little variance when changing the physical parameters inside one phase. We observe this feature also in the training of the VQAD.

\section{Results}

\subsection{\label{sec:sim} Simulations with ideal quantum data}

In order to test the performance of VQAD, we first study the one-dimensional extended Bose Hubbard model with dimerized hoppings (DEBHM) \cite{sugimoto19},
 \begin{eqnarray}\label{eq:ham_bh}
 H&=&-\sum_{i=1}^{L-1}(J+\delta J(-1)^i) (b_i^\dagger b_{i+1} + \text{h.c.})+ \nonumber \\
 &&+\frac{U}{2}\sum_i^L n_i(n_i-1)+V\sum_{i}^{L-1} n_i n_{i+1},
\end{eqnarray} 
where $b^\dagger_i(b_i)$ is the bosonic operator representing the creation(annihilation) of a particle at site $i$ of a lattice of length $L$. The tunneling amplitudes $J-\delta J$ $(J+\delta J)$ indicate hopping processes on odd (even) links connecting nearest-neighbor sites, while $V$ represents the nearest-neighbor (NN) repulsion. Here, we take the hardcore boson limit, i.e. the on-site repulsion $U/J \rightarrow \infty$, such that the local Hilbert space is two-dimensional and each site can only accommodate $0$ or $1$ bosons. This model can be effectively mapped into a spin-$1/2$ system \cite{tao13}.

Previous studies of the DEBHM model at half filling ($\bar{n} = 0.5$) have demonstrated the existence of three distinct phases \cite{sugimoto19}. For small and intermediate values of $V/J$ and $\delta J>0$, we find a topological Mott insulator (TMI) displaying features analogous to a symmetry protected topological phase appearing in the dimerized spin-1/2 bond-alternating Heisenberg model \cite{tao13}. On the other hand, for negative values of $\delta J$ we expect a trivial Mott insulator (MI), while in the regime where the nearest-neighbor repulsion dominates, a charge density wave (CDW) appears.

In \cref{fig:bh}(a)-(b), we study the phase diagram of the model in Eq.~\eqref{eq:ham_bh} in terms of the parameters $\delta J$ and $V/J$, using the density matrix renormalization group algorithm (DMRG) \cite{schollwock11,white92, tenpy}. In order to differentiate between the Mott insulating phases and the CDW, one can compute the CDW order parameter
\begin{eqnarray}\label{eq:OCDW}
O_{CDW} = \sum_{i=1}^{L/2}(-1)^{i} \delta n_i,
\end{eqnarray} 
which detects staggered patterns in the density. In \cref{fig:bh}(a) we report a vanishing value of $O_{CDW}$ everywhere but in the region with large values of $V/J$, which corresponds to the CDW \footnote{In the definition of $O_{CDW}$, we consider only half of the sites of the system because the DMRG algorithm outputs a symmetric state, which is a superposition of the two degenerate ground states.}. To characterize the TMI we study the entanglement spectrum (ES), which is expected to be doubly degenerate in a topologically non-trivial phase \cite{pollmann10} due to the existence of edge states. The entanglement spectrum $\{\lambda_i\}$ is defined in terms of the positive real-valued Schmidt coefficients $\{\alpha_i\}$ of a bipartite decomposition of the system by $\alpha_i^2 = \exp(-\lambda_i)$. We determine its degeneracy using
\begin{equation}\label{D_ES}
    D_{ES} = \sum_{i} (-1)^i e^{-\lambda_i}.
\end{equation}
In \cref{fig:bh}(b), we show that the quantity $D_{ES}$ vanishes only for small NN interaction strengths $V$ and positive values of $\delta J$, which correponds to the TMI. The trivial MI and CDW phases do not show a degeneracy and hence do not host topological edge states.
%

In the following, we test the capabilities of the VQAD with ideal states obtained from DMRG simulations. The anomaly syndrome is trained using a single representative ground state within one of the phases such that the cost measured at the trash qubits is minimised and the states of this phase can be efficiently compressed by the circuit. Afterwards, the trained circuit processes all states from the full phase diagram, ideally with similarly low cost in the same phase and significantly higher cost in other phases.

In \cref{fig:bh}(c)-(e) we show the resultant cost diagram for three circuits, each optimized at a different point in the phase diagram. Indeed, ground states outside of the training phase give rise to a large cost and hence are correctly classified by the VQAD as anomalous. Surprisingly, a single ground state example (indicated by the cross) was sufficient to successfully train the VQAD and infer all three phases. Similar results were recently reported for the case of classical anomaly detection using neural network auto-encoders \cite{Kottmann2021}.

\begin{figure}
    \centering
    \includegraphics[width=.35\textwidth]{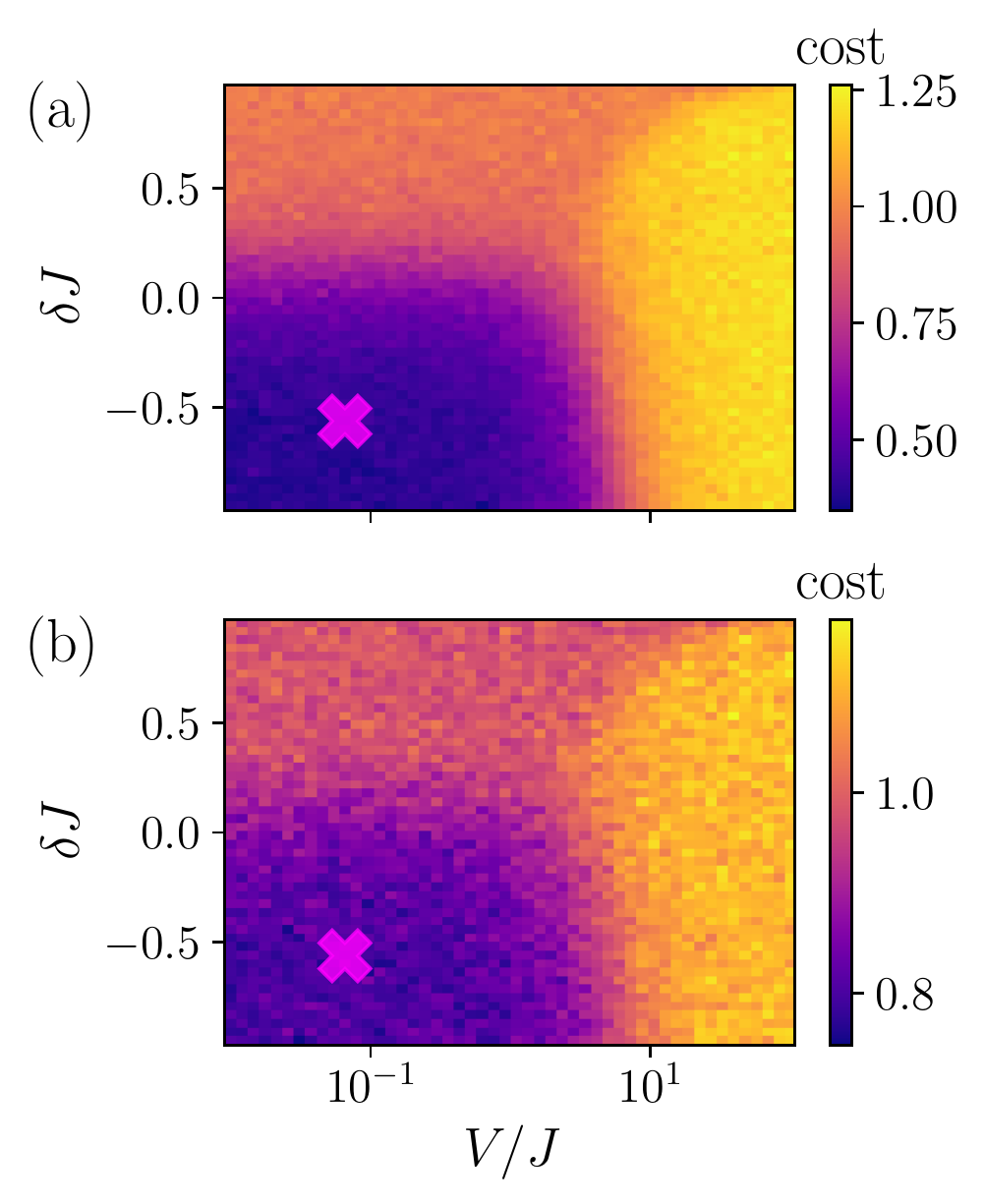}
    \caption{Cost of a VQAD trained on a single ground state in the MI phase (marked by the cross) of the DEBHM with $L=12$ sites and $n_t=2$ trash qubits. The gates of the VQAD circuit are subject to depolarizing noise with $p_{\text{err}}=0.001$ (single-qubit gates) and (a) $p_{\text{err}}=0.01$, (b) $p_{\text{err}}=0.07$ (two-qubit gates). The chosen values are motivated by the error probabilities of real devices.}
    \label{fig:bh_noisy}
\end{figure}

To demonstrate the robustness of the VQAD against noise present in currently available NISQ devices we apply a depolarizing noise channel after each gate with error probabilities $p_{\text{err}}=0.001$ (single-qubit gates) and $p_{\text{err}}=0.01,0.07$ (two-qubit gates) and show two exemplary cost profiles of the trained anomaly detector in \cref{fig:bh_noisy}. Since the noise becomes more prominent with larger circuit depths, we used the two-layer VQAD circuit ansatz with only two trash qubits in this case. While it is not possible to reach a cost of zero in the training phase, the optimization still converges and all three phases can be successfully inferred. Hence, this suggests that even if the VQAD is not able to fully disentangle the trash qubits, the phase diagram can still be recovered from the resultant cost profile.

\subsection{\label{sec:exp} Experiments on a real quantum computer}

We have seen that with ideal quantum data, VQAD can map out non-trivial phase diagrams including topologically non-trivial phases with and without noise in the anomaly syndrome. Next, we discuss its performance in real-noise simulations, that is with noise profiles and qubit connectivities from a real quantum device. Furthermore, we perform the quantum simulation subroutine, i.e., the ground state preparation via VQE, on the same circuit.
For this task, we consider the paradigmatic transverse longitudinal field Ising (TLFI) model \cite{fogedby78}
\begin{equation}
	H=J\sum_{i=1}^{L} Z_{i} Z_{i+1}-g_{x} \sum_{i=1}^{L} X_{i}-g_{z} \sum_{i=1}^{L} Z_{i},
	\label{IsingHam}
\end{equation}
where $X_{i},Z_{i}$ are the Pauli matrices on site $i$, $J$ is the coupling strength, and $g_x, g_z$ are the transverse and longitudinal fields, respectively. For $g_z=0$ the model is exactly solvable and shows a quantum phase transition from a ferromagnetic (antiferromagnetic) phase for $g_x/J<1$ and $J$ negative (positive) to a paramagnetic one for $g_x/J>1$ \cite{sachdevqpt}. In the following we set $J=1$ and vary the longitudinal and transverse fields. In this regime the model is not exactly solvable and the phase diagram has been extensively studied numerically \cite{sen00,ovchinnikov03}. The antiferromagnet-paramagnet quantum phase transition is best characterized by the order parameter which in this case is the staggered magnetization
\begin{equation}
\hat{S}=\sum_{i=1}^L (-1)^i \frac{Z_i}{L}.
\end{equation}

We simulate the ground states of the Hamiltonian in \cref{IsingHam} using VQE for $L=5$. On a noisy device, long-range entangling gates are performed by consecutive local two-qubit gates (SWAP operation), increasing the actual circuit depth. A large number of consecutive gates leads to decoherence due to gate errors and destroys the results. With the circuit presented in \cref{fig:overview} for the VQE subroutine, we found a trade-off between expressibility and noise tolerance with a circular entanglement distribution and only one layer. Additionally, we performed measurement error mitigation \cite{Bravyi2021}, which can further improve the results of the cost function as seen in \cref{fig:mitigation} in App.~\ref{app}.

\begin{figure}
    \centering
    \includegraphics[width=.35\textwidth]{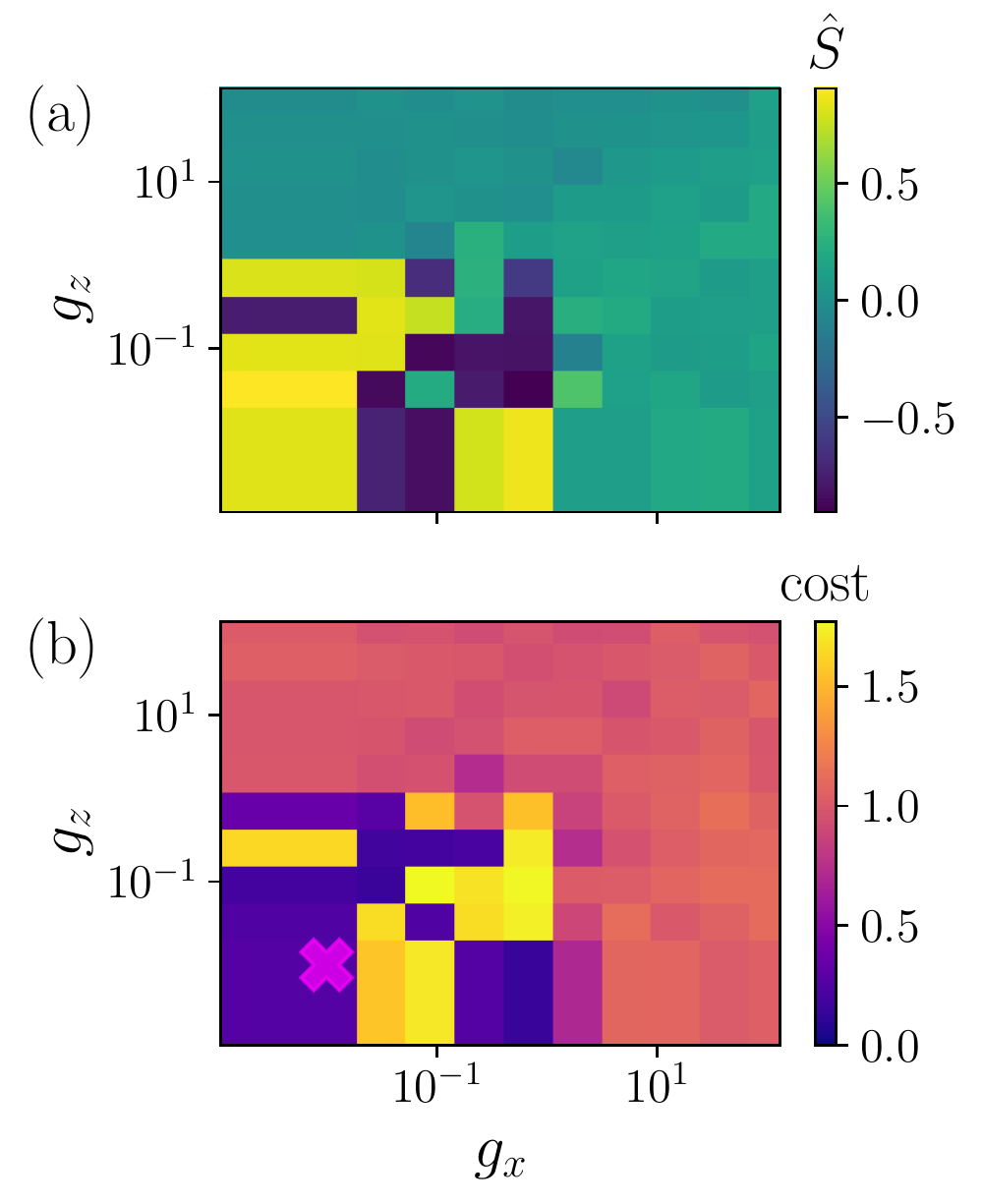}
    \caption{Real-noise simulations of the staggered magnetization $\hat{S}$ (a) and the anomaly syndrome (b) for the TLFI model. We trained the anomaly syndrome in the ordered phase on a state with positive $\hat{S}$, indicated by the purple cross. Inside the ordered phase, there is a perfect correlation between low cost states for positive $\hat{S}$, and very high cost where VQE converged to a negative $\hat{S}$. The paramagnetic phase is detected by a plateau in the anomaly syndrome.}
    \label{fig:antiferro2D}
\end{figure}

For small values of $g_x$ and $g_z$, in the ferromagnetic ordered phase, the ground states $\psi\simeq\ket{10101}$ ($\langle\hat{S}\rangle=1$) and $\psi\simeq\ket{01010}$ ($\langle\hat{S}\rangle=-1$) have a similar energy, which is why the optimization can get stuck in local minima. Hence, in the ordered phase, VQE can converge to both a state with positive or negative staggered magnetization, or an equal superposition of the two as can be seen in \cref{fig:antiferro2D}(a). The VQAD simulation results in \cref{fig:antiferro2D}(b) show a perfect correlation between positive $\braket{\hat{S}}$ and low cost, and vice versa, negative $\braket{\hat{S}}$ and high cost - which, intuitively,  can be expected \footnote{In a very hand-wavy way, we can understand this as we train the circuit $U$ to perform $U \ket{10101} = \ket{\Psi} \otimes \ket{00}_\text{trash}$ such that $U \ket{01010} = \ket{\Psi} \otimes \ket{11}_\text{trash}$ if we input a state with opposite ordering.}. The disordered phase is detected from the plateau of high cost ($\sim 1$).

\begin{figure}[h]
    \centering
    \includegraphics[width=\columnwidth]{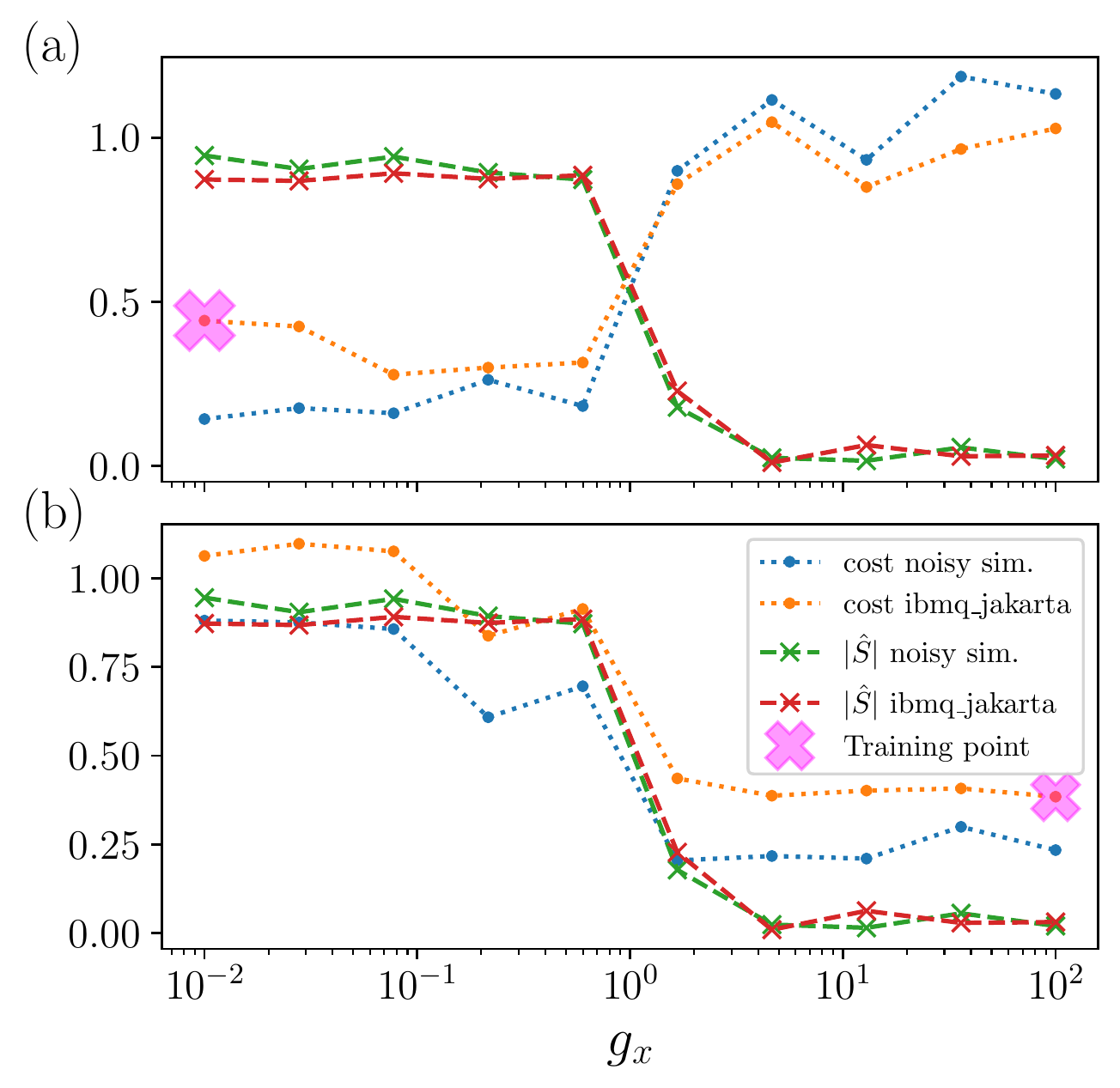}
    \caption{Real device VQAD experiments:
    We show the order parameter $\hat{S}$ compared to the VQAD results both for execution on \code{ibmq_jakarta} and noisy simulators with the same noise profile. We trained on a single ground state in the ordered (a) and paramagnetic (b) phase. 
    For sampling $\hat{S}$, we use the same parameters for the VQE circuit in simulation and experiment. All values for $\hat{S}$ in the paramagnetic phase are negative, hence, for better visualization we plot its absolute value $|\hat{S}|$.
    For training the anomaly syndrome, the optimized parameters from the simulation are taken as an initial guess.
    }
    \label{fig:ibmq_main}
\end{figure}

We see that VQAD also performs well under realistic conditions, so we next test the algorithm on a physical device. For this task, we use the $L=5$ qubits on \code{ibmq_jakarta} \cite{Bravyi2021}. To avoid jumps in the staggered magnetization in the ordered phase and improve convergence of the VQE optimization, we reuse already optimized parameters at neighboring points in the phase diagram as a good initial guess. Due to a large computation time overhead per execution on the real device, we additionally prepared pre-optimized parameters for both subroutines from a realistic noisy simulation, and use these as initial guesses for the optimization on the device. We found that for computing the staggered magnetization it is actually not necessary to re-run the VQE optimization on the physical device, and we can achieve faithful results by directly using the optimized parameters from the simulation as seen in \cref{fig:ibmq_main}. The resulting cost values for the optimized circuit, plotted in \cref{fig:ibmq_main}, clearly distinguish the two phases, with the cost from the experiment showing solely an almost constant offset compared to the noisy simulation.

\section{Outlook}

We showed that our proposed algorithm is capable of mapping out complex phase diagrams, including topologically non-trivial phases. We further demonstrated that the algorithm also works in realistic scenarios for both real-noise simulations and on a real quantum computer. Hence, we provide a tool to experimentally explore phase diagrams in future quantum devices, which will be especially useful when physical devices surpass the limit of what can be classically computed.

Currently, the main bottleneck of VQAD is the presence of noise in real devices. We were able to improve our anomaly detection scheme by employing measurement error mitigation and adopting the circuits according to the physical device. These results are promising, and with current efforts on enhancing device performances, error mitigation and circuit optimization strategies in the community, we are hopeful to see even further improvements soon.

In this work we focused on using VQAD to extract the phase diagram of quantum many-body systems. A possible future extension would be to apply it to the problem of entanglement witnessing and certification in many-body scenarios without tomography. Furthermore, the use of an autoencoder-like architecture has the advantage over kernel-based schemes in that there exists tools of interpreting the feature space in classical autoencoders to gain physical insights \cite{Iten2018}, which can be a possible future extension for the quantum case discussed here.

\begin{acknowledgments}
\paragraph*{Acknowledgments}
The authors thank D. Gonz\'alez-Cuadra, L. Barbiero and U. Bhattacharya for discussions on the extended Bose Hubbard model and A. Dauphin, J. Bowles and M. Lewenstein for valuable feedback on the manuscript. 

This work was supported by the European Union’s Horizon 2020 research and innovation programme under the Marie Sklodowska-Curie grant agreement  No 713729  (K.K.).
This work was supported by OIST Graduate University and we are grateful for the help and support provided by the Scientific Computing
and Data Analysis Section of the Research Support Division at OIST.
N.B. acknowledges support from a ``la Caixa” Foundation (ID 100010434) fellowship. The fellowship code is  LCF/BQ/DI20/11780033.
We acknowledge support from ERC AdG's NOQIA and CERQUTE, Spanish MINECO (FIDEUA PID2019-106901GB-I00/10.13039 / 501100011033, FIS2020-TRANQI, Severo Ochoa CEX2019-000910-S and Retos Quspin), the Generalitat de Catalunya (CERCA Program,  SGR 1341, SGR 1381 and QuantumCAT), Fundacio Privada Cellex and Fundacio Mir-Puig, MINECO-EU QUANTERA MAQS (funded by State Research Agency (AEI) PCI2019-111828-2 / 10.13039/501100011033), EU Horizon 2020 FET-OPEN OPTOLogic (Grant No 899794), and the National Science Centre, Poland-Symfonia Grant No. 2016/20/W/ST4/00314. 

We acknowledge the use of IBM Quantum services for this work. The views expressed are those of the authors, and do not reflect the official policy or position of IBM or the IBM Quantum team. 

\paragraph*{Author contributions}
\noindent KK and FM contributed equally to this work.

\noindent KK initiated and managed the project, performed the VQAD simulations for the TLFI model, and was in charge of the ansatz scaling and the experiments on the physical device.

\noindent FM was in charge of the implementation in Python/Qiskit, worked on the TLFI model and performed the VQAD simulations for the DEBHM.

\noindent JF was in charge of DEBHM, the DMRG simulations and performed VQE simulations for the TLFI model.

\noindent NB worked on the implementation of the TLFI model and performed VQE simulations, and worked on error mitigation in noisy simulations.

\noindent All authors contributed to the discussions and the writing of the manuscript.
\end{acknowledgments}


\bibliographystyle{unsrtnat}
\bibliography{lit.bib}

\appendix
\vfill\null
\section{\label{app} Technical details}

\begin{figure}[h!]
    \centering
    \includegraphics[width=.75\columnwidth]{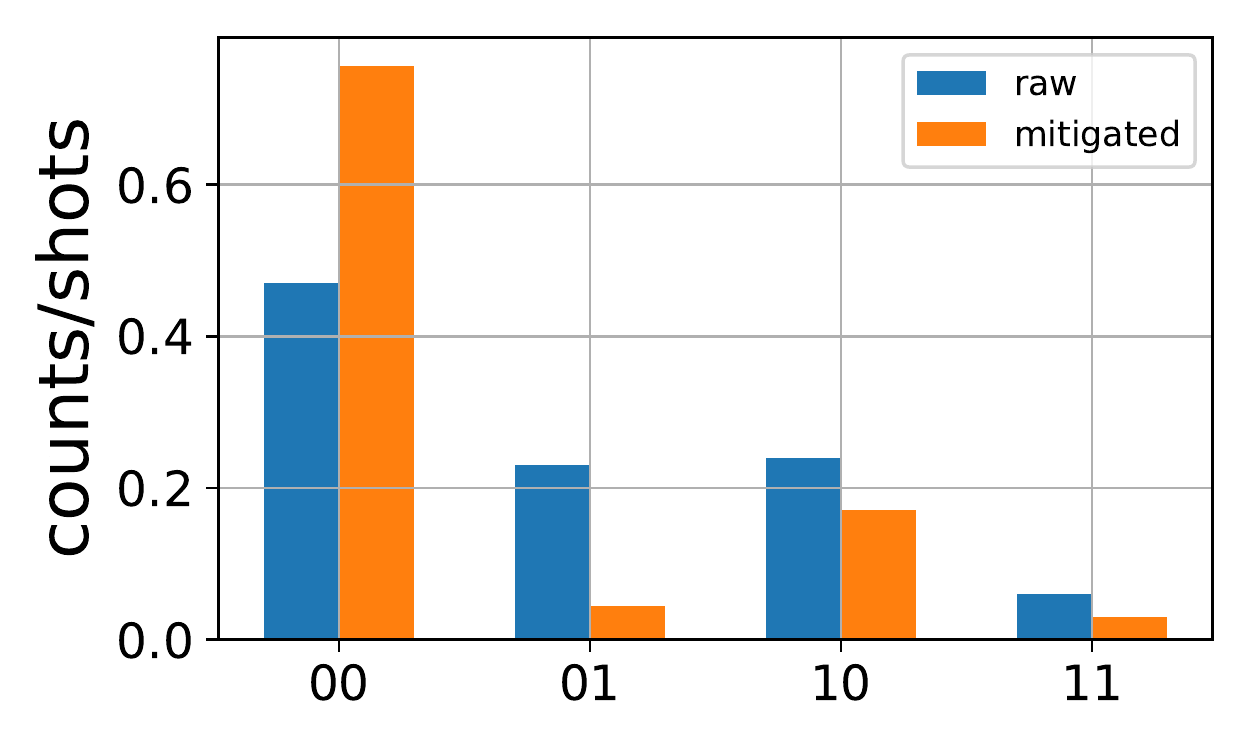}
    \caption{Comparison of the trash qubit measurement outcomes with and without measurement error mitigation. The anomaly syndrome circuit has been trained with and without error mitigation on a ground state of the TLFI model in the ordered phase in real-noise simulations. Ideally, all of the $1000$ shots would result in the $00$ bit string. By mitigating the measurement errors we improve the results towards this desired outcome.}
    \label{fig:mitigation}
\end{figure}
The code to run the simulations and experiments discussed in the main text can be found in our repository on \textsf{GitHub} \cite{github}. The optimization of the circuit parameters was performed using simultaneous perturbation stochastic approximation (SPSA) \cite{Spall1998,Kandala2017}.
To obtain the results presented in \cref{fig:bh} of Sec.~\ref{sec:sim}, a VQAD circuit ansatz composed of 6 layers (6 trash qubits) was employed resulting in $6L+6$ parameters. For the noisy simulations and real-device execution discussed in Sec.~\ref{sec:exp}, we used the ansatz in \cref{fig:overview}, counting $2L$ and $2L+2$ parameters for the quantum simulation and anomaly syndrome, respectively. In classical real-noise simulations, we used $500$ VQE optimization iterations for the initial ground state optimization, and $200$ iterations for all subsequent optimizations where the previously optimized parameters were taken as initial guesses. For the anomaly detection circuit, we found converged results with less than $100$ optimization iterations. As an example, calculating the expectation value of the magnetization takes roughly $2-10$ seconds on a commercial laptop (here: i7-4712HQ), while the real-device execution takes about $30$ seconds. Furthermore, we used measurement error mitigation \cite{Bravyi2021} provided by the Qiskit library to improve the results of the VQAD simulations in the presence of noise as illustrated in \cref{fig:mitigation}.

\end{document}